\documentclass[aps, twocolumn, groupedaddress, prx]{revtex4-2}
\usepackage{graphicx}
\usepackage{amsmath}
\usepackage[colorlinks=true, linkcolor=blue, citecolor=blue, urlcolor=blue]{hyperref}
\begin{document}

\preprint{APS/123-QED}

\title{A quantitative approach to flowing supercooled liquids: From microscopic heterogeneities to rheology}

\author{Dong-Xu Yu and Zhe Wang}
\email{zwang2017@mail.tsinghua.edu.cn}
\affiliation{Department of Engineering Physics and Key Laboratory of Particle and Radiation Imaging (Tsinghua University) of Ministry of Education, Tsinghua University, Beijing 100084, China}

\date{\today}

\begin{abstract}

Soft glassy materials display rich and complex flow behaviors across both macroscopic and molecular scales, 
and a fundamental understanding of these phenomena remains an outstanding challenge. 
Here, we propose a theoretical model for the flow of supercooled liquids---a typical class of glassy fluids---based on a two-state paradigm 
that conceptualizes the flow as a dynamic coexistence of transient solid-like and liquid-like regions.
The model rests on two essential physical ingredients: a correlation length that captures medium-range structural order, 
and a localized elasticity-mediated interaction that restricts stress propagation within solid-like regions. 
Remarkably, with all parameters determined solely from equilibrium state, 
the model quantitatively reproduces rheological responses---including both steady-state and start-up shear---for a broad range of shear rates. 
Furthermore, it simultaneously captures the evolution of molecular dynamic heterogeneity. 
This dual success---spanning macroscopic rheology and microscopic spatiotemporal fluctuations---underscores the pivotal role of structural and 
dynamic heterogeneities in governing the rheological response. 
Moreover, it provides a direct understanding of how the flow behaviors of a supercooled liquid are embedded in its equilibrium properties.

\end{abstract}


\maketitle

\section{Introduction}

\label{sec:intro}

Flowing supercooled liquids are ubiquitous and important across a broad range of science and technology \cite{larson1999book, debenedetti1996book}. 
In life sciences, the flow of supercooled liquids underlies various transport processes in living systems exposed to subfreezing conditions \cite{supercoolanimal2015,debenedetti1996book}; 
and in industrial processing, precisely controlling the flow of supercooled melts plays a key role in growth of crystals \cite{crystal_growth_2018}, 
metal molding \cite{schroers_2010_advmat}, and thermoplastic forming of glasses \cite{liu_2016_jom, baran2022pmatsci}. 
Flowing supercooled liquids exhibit remarkable diversity---both in the mechanical response to external deformation, and in the spatiotemporal organization of molecular motion. 
As for the former, typical examples include shear thinning under steady shear and stress overshoot in start-up shear \cite{yamamoto1997epl, tanaka2009prl, petekidis2012prl, schall2016sofmat}. 
As for the latter, the molecular relaxation in flow exhibits heterogeneous distributions in space, known as dynamic heterogeneity (DH) \cite{yamamoto1998pre,yamamoto2012jcp}. 
These phenomena sensitively depends on the flow rate and the degree of supercooling. 

The understanding of the origin of this diversity remains a major challenge in the liquid-state physics and is in a state of flux. 
Works by Yamamoto and Onuki identify a connection between shear thinning and the shrinkage of bond-breaking clusters, 
which implies the central role of DH in governing the flow behaviors \cite{yamamoto1998pre,yamamoto1997epl}. 
In a similar spirit, the random first-order transition theory attributes shear thinning to cooperative relaxations over finite regions \cite{rfotshearthinning2009pnas}. 
By contrast, the mode-coupling theory and some other approaches emphasize the cage-scale dynamics \cite{mctshearthinning2008prl, miyazaki2024comphy,furukawa2023prr,fuchs2013jor}. 
In these studies, nonlinear rheology is attributed to local relaxations via advection-assisted cage distortion and escape, whereas the role of DH is obscure. 
Besides this inconsistency, another problem in current theoretical literature is the much less attention paid to the transient response. 
As mentioned above, supercooled liquids exhibit strong stress overshoot in response to the start-up shear. 
This behavior is not only a common nonlinear phenomenon of great interest \cite{benzi2021prl,bonn2017rmp,egelhaaf2017prl,voigtmann2020cps}, 
but also the key process for bridging the flow behaviors to the equilibrium properties of liquids \cite{fuchs2008jpcm,fuchs2013jor}. 
There is, thus, a strong need to seek a unified picture for describing the steady-state rheology, the transient mechanical response, the evolution of DH, 
and how they are determined by the quiescent properties of supercooled liquids.

The ultra-high viscosity of supercooled liquids can be understood by envisioning them as solids that flow \cite{dyre2024jpcl, dyre2006rmp, lemaitre2013prl,dyre1999pre,egami2012prl}, 
in the sense that they transiently behave elastically below certain length scales. 
Inspired by this idea, we proposed a concept of localized elastic region (LER) to explain the shear thinning of supercooled liquids \cite{wang2022prx,wang2025jcp}. 
LERs are short-lived, mesoscopic regions, which deform elastically before yielding, providing the resistant stress to imposed shear. 
The LER picture implies that within the flow there exist two types of response, the solid-like response and the liquid-like response \cite{wang2025jcp,wang2025prr}. 
Building on this \emph{two-state} idea, herein, we establish a model for flowing supercooled liquids by introducing two critical conceptual ingredients: 
(i) a correlation length that characterizes the medium-range reorganization of local packing, and 
(ii) the localization of the elasticity-mediated interaction. 
The resulting model quantitatively predicts the shear-rate dependence of viscosity under steady shear and the transient stress response during start-up shear. 
It also captures the shear-rate-dependent evolution of DH. 
Notably, all model parameters are determined from the quiescent properties, enabling predictions without fitting to flow data. 

\section{Background}

\subsection{Flow Behaviors of Supercooled Liquids}

\label{sec:md_simulation}

\begin{figure}
	\includegraphics[width=\linewidth]{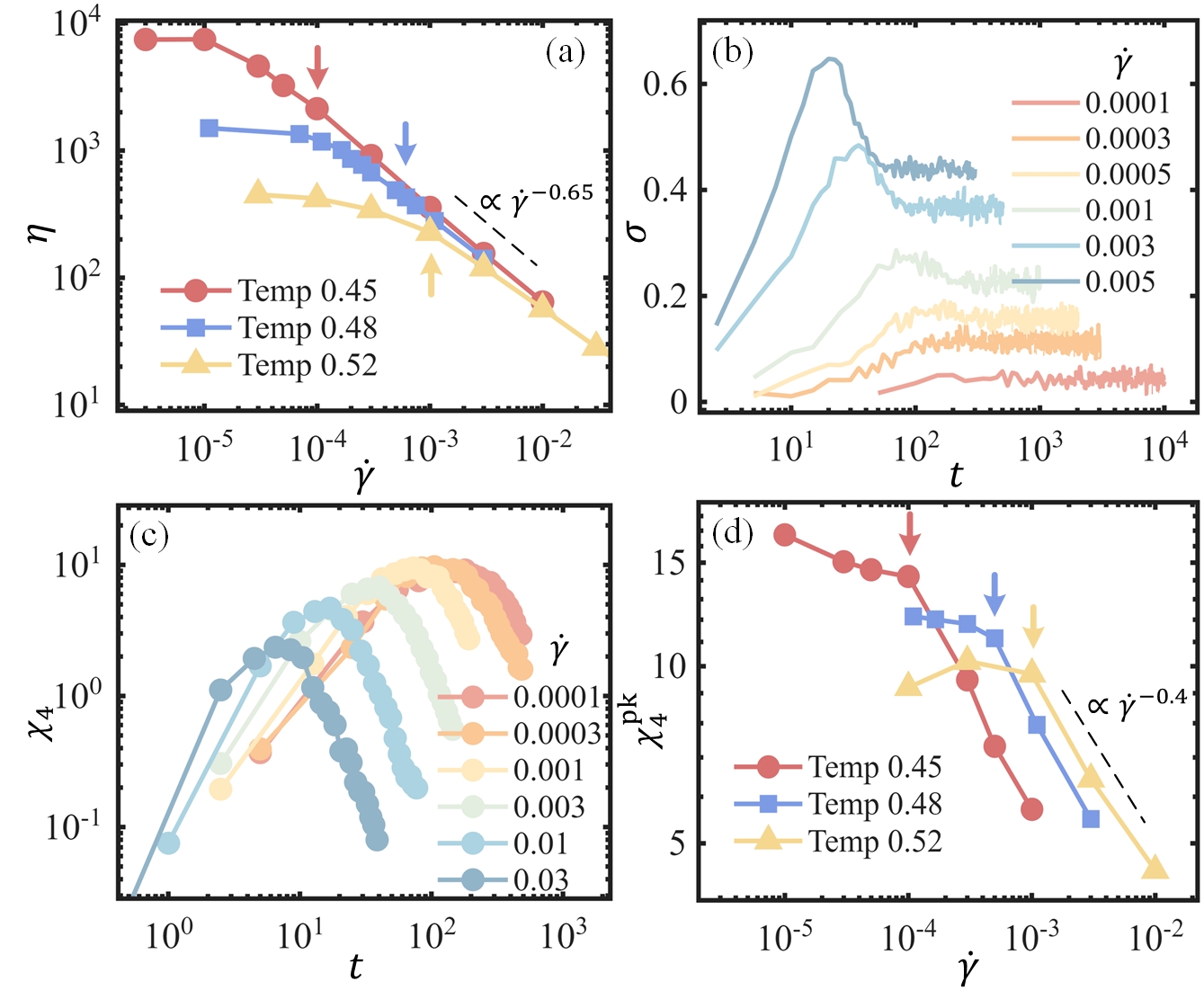}
	\caption{\label{fig:md_info}
		Flow behaviors of supercooled liquids (KA system). 
		(a) Viscosity $\eta$ as a function of shear rate $\dot{\gamma}$ under steady shear for three temperatures. 
			Arrows indicate the onset $\dot{\gamma}$ of the power-law-thinning $\dot{\gamma}_\mathrm{pl}$. 
		(b) Start-up stress $\sigma(t)$ at $T=0.52$ for various $\dot{\gamma}$. 
		(c) Four-point susceptibility $\chi_4(t)$ at $T=0.52$ for various $\dot{\gamma}$. 
		(d) Peak values $\chi_4^\mathrm{pk}$ versus $\dot{\gamma}$ for three temperatures, where arrows mark $\dot{\gamma}_\mathrm{pl}$. 
	}
\end{figure}

We first review some representative phenomena of flowing supercooled liquids based on 
the molecular dynamics (MD) simulation of the Kob-Andersen (KA) mixture of Lennard-Jones (LJ) particles \cite{kobandersenmixture} (see Appendix~\ref{sec:appendix_sim}). 
Figure~\ref{fig:md_info}(a) shows the viscosity $\eta$ as a function of shear rate $\dot{\gamma}$ under steady shear at different temperatures $T$. 
For all cases, $\eta(\dot{\gamma})$ remains Newtonian at low $\dot{\gamma}$, and exhibits a thinning following $\eta(\dot{\gamma}) \sim \dot{\gamma}^{-\lambda}$ 
with $\lambda \approx 0.65$ at high $\dot{\gamma}$ \cite{barrat2002jcp,miyazaki2024comphy,furukawa2023prr}. 
We denote the $\dot{\gamma}$ at which $\eta(\dot{\gamma})$ enters the power-law-thinning regime as $\dot{\gamma}_\mathrm{pl}$, 
and mark them in Fig.~\ref{fig:md_info}(a) by arrows. 
Between the Newtonian regime and the power-law-thinning regime there is a crossover regime \cite{wang2025prr,furukawa2017pre}, as seen in Fig.~\ref{fig:md_info}(a). 
Figure~\ref{fig:md_info}(b) shows the transient responses to start-up shear. 
As $\dot{\gamma}$ increases, a progressively pronounced stress overshoot emerges. 

At the microscopic level, the DH can be characterized by the four-point susceptibility $\chi_4(t)$ \cite{glotzer2003jcp} adapted to the flow condition \cite{yamamoto2012jcp}: 
$\chi_4(t)=N\left[\langle Q^2(t)\rangle-\langle Q(t)\rangle^2\right]$, 
where $Q(t)=\sum_{i=1}^{N} H\!\left(a-\left|\delta \vec{r}_i(t)\right|\right)/N$,
$H$ is the Heaviside function, $\delta \vec{r}_i(t)$ is the nonaffine displacement of particle $i$ over time $t$ \cite{yamamoto1998pre}, and $a$ is set to $0.3$. 
Figure~\ref{fig:md_info}(c) shows $\chi_4(t)$ at $T=0.52$ for different $\dot{\gamma}$. 
In all cases, $\chi_4(t)$ exhibits a peak $\chi_4^\mathrm{pk}$, which quantifies the strength of DH \cite{glotzer2003jcp}. 
The $\dot{\gamma}$-dependence of $\chi_4^\mathrm{pk}$ is summarized in Fig.~\ref{fig:md_info}(d). 
A clear crossover is observed: as $\dot{\gamma}$ increases, $\chi_4^\mathrm{pk}$ crosses over to a power-law shrinkage $\chi_4^\mathrm{pk} \sim \dot{\gamma}^{-0.4}$. 
We mark the crossover shear rates with arrows in Fig.~\ref{fig:md_info}(d). 
Interestingly, they coincide $\dot{\gamma}_\mathrm{pl}$. 
Similar trends have been reported previously \cite{yamamoto2012jcp}. 

\subsection{Two-State View of Flowing Supercooled Liquids}

\label{sec:ler_review}

\begin{figure}
	\includegraphics[width=\linewidth]{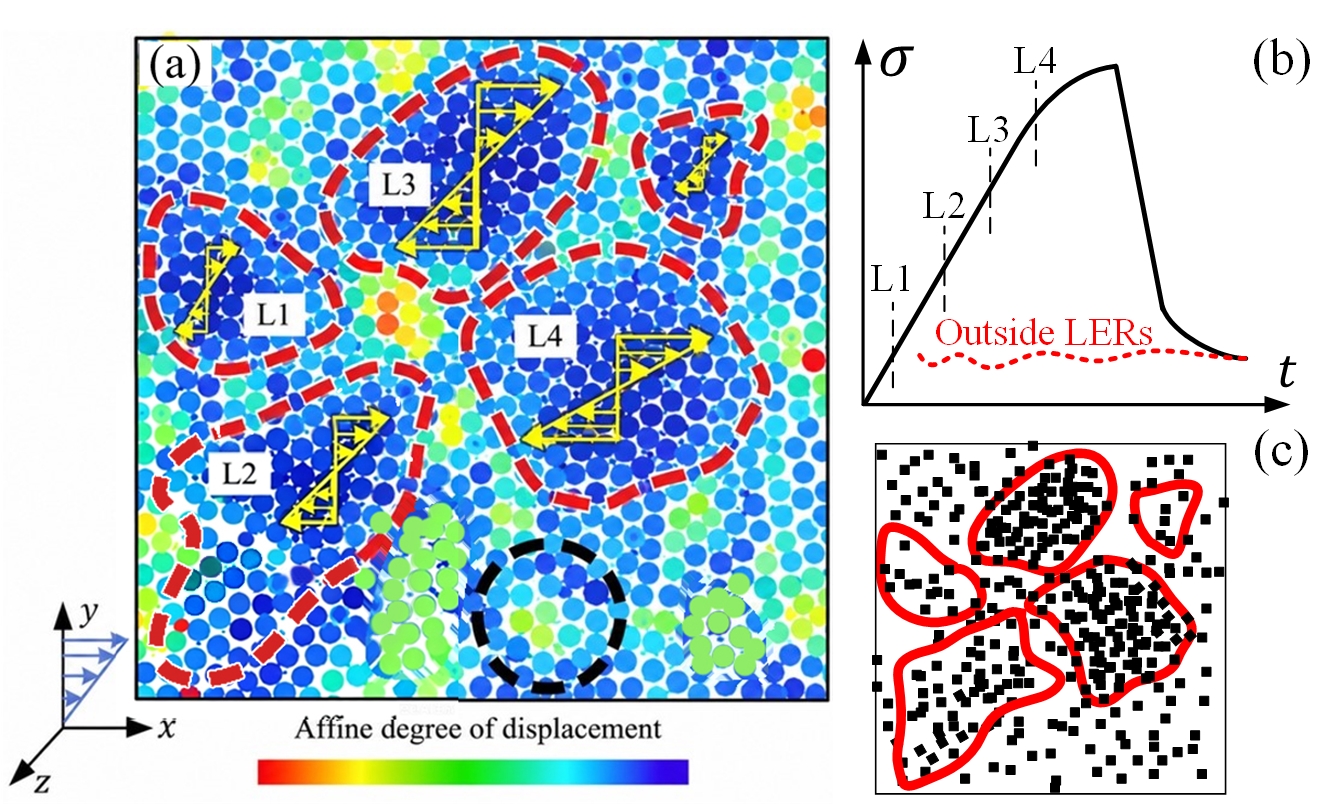}
	\caption{\label{fig:ler_review}
		Illustration of localized elastic regions (LERs) in a flowing supercooled liquid. 
		(a) Snapshot of a sheared configuration with particles colored by the affine degree of their displacements over a time interval. 
			Four representative LERs, labeled L1--L4, are outlined with red dashed lines, and yellow arrows inside each LER indicate the magnitude of its elastic deformation. 
			A region lying outside any LER is outlined with a black dashed line for comparison. 
		(b) Stress evolution inside an LER (black solid line) and outside LERs (red dashed line). 
			LERs accumulate stress elastically before yielding, whereas regions outside LERs exhibit no appreciable stress buildup. 
			The four LERs in (a) lie at different stages of elastic loading, as marked by the vertical dashed lines. 
		(c) Relaxation events (black squares) accumulated over a structural relaxation time $\tau_\alpha$. 
			Their spatial clustering, a feature of DH, coincides with regions previously occupied by highly deformed LERs (red outlines). 
		See Ref.~\cite{wang2025jcp} for a full account of the LER picture.
	}
\end{figure}

Why supercooled liquids, nominally in liquid state, exhibit progressively stronger nonlinear viscoelasticity as $\dot{\gamma}$ increases?
The LER picture \cite{wang2022prx,wang2025jcp,wang2025prr} provides an answer. 
As illustrated in Fig.~\ref{fig:ler_review}(a), a flowing supercooled liquid is not uniform: 
It contains transient solid-like regions, i.e., LERs, embedded in the liquid background \cite{wang2022prx}. 
Each LER contains hundreds of particles that undergo coherent affine displacements in flow \cite{wang2025jcp}. 
This affinity enables the solid-like stress accumulation within LER, followed by plastic yielding and rearrangement (Fig.~\ref{fig:ler_review}(b)). 
By contrast, regions outside LERs exhibit viscous behavior with no appreciable stress buildup, a hallmark of normal liquid response \cite{hess1986pra} (Fig.~\ref{fig:ler_review}(b)). 
At any given instant, the system contains multiple LERs, each at a different stage of its loading cycle. 

To explore the relation between the solid-liquid duality in local mechanical response and local dynamics, 
we introduced a particle-wise local configurational relaxation time $\tau_\mathrm{LC}$ \cite{wang2025prr}. 
$\tau_\mathrm{LC}$ is the isoconfigurational-ensemble-averaged persistence time \cite{pastore2021jcp} for a particle to undergo cage-jump \cite{biroli2010prl} from a given configuration. 
To measure $\tau_\mathrm{LC}$, one first switches off the flow at a given time, then measures the first-jump time for a reference particle, 
and finally performs the isoconfigurational average \cite{harrowell2006prl}. 
By ceasing the flow, $\tau_\mathrm{LC}$ highlights the intrinsic relaxation determined by configuration. 
Whether a region's response is solid-like or liquid-like depends on the competition between its intrinsic local mobility, represented by $\tau_\mathrm{LC}$, and the external shear rate. 
Similar to Maxwell's picture \cite{larson1999book, barrat2019prl, yip2024jcp}, 
if the intrinsic mobility is slower than the external rate, the local region exhibits solid-like response, otherwise liquid-like. 
By increasing external rate, more particles are recruited into the solid-like population, i.e., into LERs, and the rheology crosses over from linear to nonlinear \cite{wang2025prr}. 

As LER' strain reaches the yielding point, particles inside the LER collectively rearrange, resulting in a dynamic cluster. 
In this way, the emergence of prominent DH is linked to the massive yielding of LERs \cite{wang2025jcp}, as illustrated in Fig.~\ref{fig:ler_review}(c). 

\section{Two-State Model}

\begin{figure}
	\includegraphics[width=\linewidth]{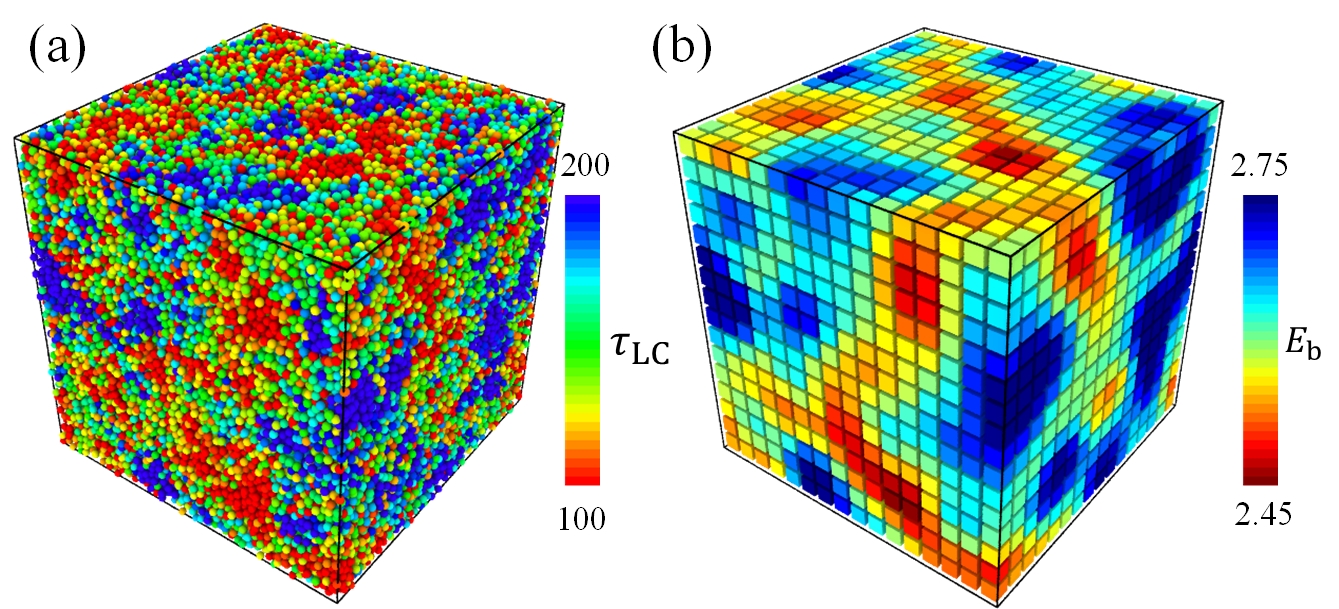}
	\caption{\label{fig:taulc_eb}
		Illustration of the spatial discretization. 
		(a) Spatial distribution of $\tau_\mathrm{LC}$ in the MD configuration of KA system at $T=0.52$. 
		(b) Corresponding coarse-grained energy barrier field $E_\mathrm{b}$ mapped onto blocks.
	}
\end{figure} 

In equilibrium state, $\tau_\mathrm{LC}$ is shown to define a local energy barrier $E_\mathrm{b}$, 
which is determined by local configuration \cite{ajliu2016nat}, through the activation picture \cite{wang2025prr}:
\begin{equation}
	\label{eq:taulc_eb}
	\tau_\mathrm{LC}\sim \exp\!\left(\frac{E_\mathrm{b}}{k_\mathrm{B}T}\right),
\end{equation}
where $k_\mathrm{B}$ is the Boltzmann constant. 
Thus, by measuring the equilibrium distribution of $\tau_\mathrm{LC}$, one obtains the energy barrier density $\rho(E_\mathrm{b})$ via Eq.~\ref{eq:taulc_eb}. 
Figure~\ref{fig:taulc_eb}(a) shows an example of the spatial distribution of $\tau_\mathrm{LC}$. 
$\tau_\mathrm{LC}$, and equivalently $E_\mathrm{b}$, exhibits significant clustering in space, a typical feature of DH \cite{szamel2014prl,ediger2000review}. 
Since $E_\mathrm{b}$ varies smoothly in space, the space can be discretized into blocks with each block associated with a mean $E_\mathrm{b}$ averaged within the block \cite{martens2021prl,biroli2022prl}, 
as illustrated in Fig.~\ref{fig:taulc_eb}(b). 
The block size is determined from equilibrium spatial correlations of relaxation events \cite{chandler2011prx}, as detailed in Supplemental Materials (SM). 

Shear can facilitate the activation of block \cite{lacks2001prl}. 
For simplicity, we assume a harmonic form for the energy accumulated during the deformation of a block \cite{cates1997prl, fielding2024prl}. 
Thus, block $i$ activates with the rate
\begin{equation}
	\label{eq:activation}
	\Gamma \exp\!\left[
	-\frac{E_{\mathrm{b},i}-\sigma_i^2 \Omega /(2\mu_i)}
	{k_\mathrm{B}T}
	\right],
\end{equation}
where $\sigma_i$ is the stress of block $i$, $\mu_i$ is the shear modulus of block $i$, $\Omega$ is the activation volume, taken to be the block volume, 
and $\Gamma$ is an attempt frequency estimated from the time scale preceding the plateau in the equilibrium mean-squared displacement (MSD) \cite{cates1997prl}. 
Once activated, a block relaxes while continuing to deform, 
and the relaxation terminates when the total strain accumulated locally during this phase exceeds a restructuring strain $\gamma_\mathrm{res}$ \cite{martens2016prl}. 
The total strain rate has two contributions: an elastic part proportional to the stress rate, and a plastic part proportional to the stress relaxation rate. 
Integrating their sum gives the restructuring criterion: 
\begin{equation}
	\label{eq:restructure}
	\int \left|
	\frac{\dot{\sigma}_i}{\mu_i}
	+\frac{n_i \sigma_i}{\mu_i \tau}
	\right| \mathrm{d}t
	> \gamma_\mathrm{res},
\end{equation}
where the integration is over the relaxing phase,
$\tau$ is the characteristic relaxation time of block and is set to $\Gamma^{-1}$, 
and $n_i$ is a flag variable for block $i$: $n_i=0$ denotes the loading phase, while $n_i=1$ denotes the relaxing phase \cite{barrat2018rmp}. 
$\gamma_\mathrm{res}$ is found from a mean-field analysis detailed in SM.

The forms of Eqs.~\ref{eq:taulc_eb}--\ref{eq:restructure}, as well as the space discretization, 
can be found in some influential generic models of glassy dynamics \cite{cates1997prl,barrat2018rmp,bouchaud2003prl}.
In following parts, we will introduce several mechanisms and concepts, which highlight the distinct features of flowing supercooled liquids.

\subsection{Block Renewal and Structural Correlation Length}

\label{sec:corr_length}

After activation and restructuring, a block will start a new cycle of deformation with a new $E_\mathrm{b}$. 
The renewal of $E_\mathrm{b}$ in steady state must preserve three properties: 
steadiness of the statistical distribution of $E_\mathrm{b}$, persistence of the spatial clustering of $E_\mathrm{b}$, 
and gradual renewal of the spatial distribution of $E_\mathrm{b}$ over time \cite{castillo2023prl}. 
To fulfill these requirements, we need to figure out what factors affect the renewal of $E_\mathrm{b}$ of a block. 
First, the new $E_\mathrm{b}$ must contain a component $E_\mathrm{rand}$ that represents the thermal randomness. 
$E_\mathrm{rand}$ is directly sampled from the equilibrium $\rho(E_\mathrm{b})$, as suggested by the soft glassy rheology model \cite{cates1997prl}. 
Second, the renewal of a block should be influenced by nearby blocks. 
We denote this component as $E_\mathrm{near}$, and express it as:
\begin{equation}
	\label{eq:eneigh}
	E_\mathrm{near}=\frac{1}{Z}\sum_{j} \exp\!\left(-\frac{r_{ij}}{\xi}\right) E_{\mathrm{b},j},
\end{equation}
where $j$ goes through every block, $r_{ij}$ is the distance between two blocks, $Z=\sum_{j} \exp(-r_{ij}/\xi)$ is the normalization factor, 
and $\xi$ represents the correlation length between blocks. 

\begin{figure}
	\includegraphics[width=\linewidth]{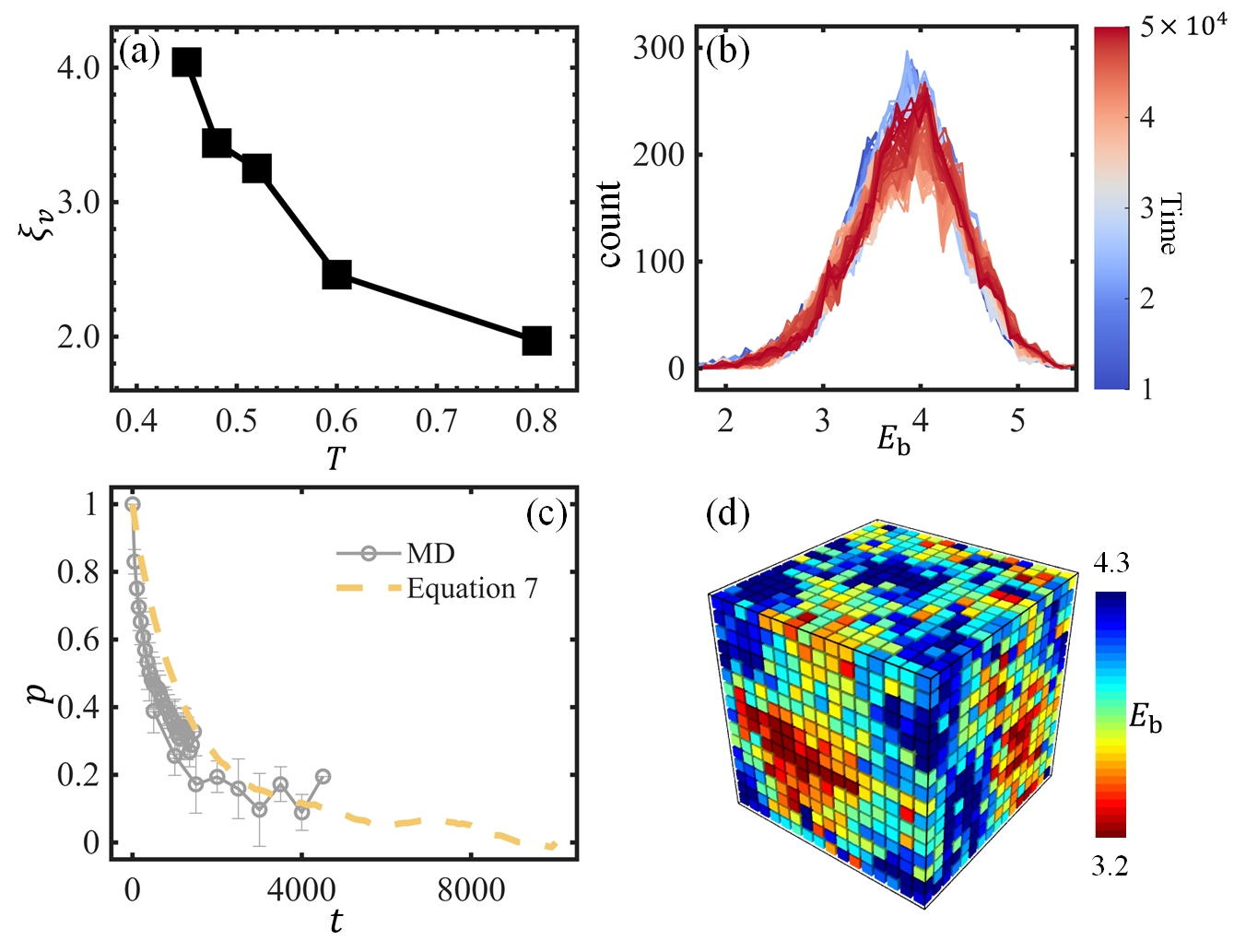}
	\caption{\label{fig:espatial_sampling}
		Renewal of $E_\mathrm{b}$ field. 
		(a) Optimal coarse-grained length $\xi_v$ as a function of $T$ for the KA system. 
		(b) Statistical distributions of $E_\mathrm{b}$ over the time window from $10^4$ to $5\times10^4$ from the model. 
		(c) Pearson correlation coefficient $p(t)$ of the $E_\mathrm{b}$ field between time $0$ and time $t$. 
			Symbols and dashed line denote MD result and model result, respectively. 
		(d) Representative spatial pattern of $E_\mathrm{b}$ from the model after a long-time evolution. 
			Clustered feature is clearly seen. 
	}
\end{figure} 

The determination of $\xi$ is crucial. 
Our previous work has established a structural basis for $\tau_\mathrm{LC}$ 
by showing that the local barrier $E_\mathrm{b}$ is primarily determined by the local packing degree \cite{wang2025prr}. 
Similar ideas can be found in other studies \cite{ajliu2016nat, tanaka2019nat}. 
Thus, $E_\mathrm{near}$ should be intimately related to the local packing of nearby regions. 
To make this relation explicit, we consider an indicator of the local free volume for particle $i$ \cite{filion2021prl}:
\begin{equation}
	v_i=\frac{1}{r_\mathrm{c}^2} \sum_{j} \exp\!\left[-\frac{(r_{\mathrm{p},ij}-r_\mathrm{c})^2}{2\delta^2}\right],
\end{equation}
where $j$ goes through every particle, $r_{\mathrm{p},ij}$ is the distance between two particles, 
$r_\mathrm{c}$ is the position of the first minimum of the pair distribution function $g(r)$, and $\delta$ is set to $0.1$. 
We further coarse-grain the field of $v_i$ by:
\begin{equation}
	v_i^\mathrm{cg}=\frac{1}{Z_v} \sum_{j} \exp\!\left(-\frac{r_{\mathrm{p},ij}}{\xi_v}\right) v_j,
\end{equation}
where $Z_v=\sum_{j} \exp(-r_{\mathrm{p},ij}/\xi_v)$, $\xi_v$ is the coarse-graining length. 
As suggested by Tong and Tanaka \cite{tanaka2018prx}, the $\xi_v$ that maximizes the correlation between 
the field of a coarse-grained structural parameter ($v_i$) and the field of a particle-wise dynamic parameter ($\tau_\mathrm{LC}$) defines a structural correlation length. 
We set $\xi$ to be this optimal $\xi_v$. 
Figure~\ref{fig:espatial_sampling}(a) shows $\xi$ as a function of $T$ for the KA system. 
As expected, $\xi$ exhibits a growth as the system approaches glass transition. 

$\xi$ is determined from quiescent configurations. 
In our model, we use it without modification under shear. 
To justify this, we apply the same coarse-graining procedure to flowing configurations, 
and the results confirm that this structural length is indeed insensitive to flow rates. 

By combining $E_\mathrm{rand}$ and $E_\mathrm{near}$ to represent the renewed $E_\mathrm{b}$ in our model, 
we find that the statistical properties and the spatial clustering feature of $E_\mathrm{b}$ can be preserved, 
while the spatial distribution of $E_\mathrm{b}$ is stuck over unreasonable long time. 
To fix this problem, we empirically introduce a term, $E_\mathrm{drift}$, to account for the random migration of the spatial clustering of $E_\mathrm{b}$. 
Thus, the new $E_\mathrm{b}$ of block $i$ is given by:
\begin{equation}
	\label{eq:spatial_sampling}
	E_{\mathrm{b},i}
	= c_1 E_{\mathrm{rand}, i}
	+ c_2 E_{\mathrm{near}, i}
	+ \sqrt{1-c_1^2-c_2^2}\,E_{\mathrm{drift}, i}.
\end{equation}
The need for $E_\mathrm{drift}$ can be understood as follows. 
$E_\mathrm{near}$ at block $i$ is a weighted average of the \emph{current} $E_\mathrm{b}$ values of its neighbors, 
so updates driven by $E_\mathrm{near}$ inherit precisely the spatial pattern they are meant to refresh. 
This produces a self-reinforcing feedback: blocks in a high-$E_\mathrm{b}$ region are continually fed high values, and blocks in low-$E_\mathrm{b}$ regions remain low, 
anchoring the clusters to their initial positions. 
The $E_\mathrm{rand}$ component injects fluctuations but, being spatially uncorrelated, only blurs values locally and cannot displace clusters as a whole. 
To break this stagnation, $E_\mathrm{drift}$ is expected to provide a spatially correlated random field that is statistically independent of the existing pattern. 
Mixing it in can drive a gradual migration of the spatial pattern while preserving the statistical distribution and the clustering feature of $E_\mathrm{b}$. 
For this purpose, we construct the $E_\mathrm{drift}$ field by first establishing a discretized field with each block being randomly assigned $1$ or $-1$, 
and then convolving this random field over a range $\xi$. 
To preserve the statistical distribution of $E_\mathrm{b}$, we further rescale this field to match its variance to that of $\rho(E_\mathrm{b})$. 
The detailed instruction is given in SM. 
We tentatively set $c_1=c_2=1/\sqrt{3}$. 
The reasonableness of this choice should be validated. 

Figure~\ref{fig:espatial_sampling}(b)--(d) examine the effectiveness of Eq.~\ref{eq:spatial_sampling} for the equilibrium state at $T=0.45$. 
In Fig.~\ref{fig:espatial_sampling}(b), we show the model result of the evolution of the statistical distribution of $E_\mathrm{b}$, which is preserved in the whole testing time window. 
Figure~\ref{fig:espatial_sampling}(c) shows the Pearson correlation coefficient $p(t)$ between the $E_\mathrm{b}$ fields at the time origin and at a later time $t$, 
which quantifies how much of the original spatial pattern is retained over the interval $t$. 
The decay of $p(t)$ toward zero indicates that the spatial pattern has been effectively renewed. 
The model result and the MD result show similar decay rates, indicating that Eq.~\ref{eq:spatial_sampling} successfully captures the renewal dynamics. 
Figure~\ref{fig:espatial_sampling}(d) shows the spatial distribution of $E_\mathrm{b}$ after long-time evolution.
The clustering feature is well preserved. 
Together, the results in Fig.~\ref{fig:espatial_sampling} demonstrate that Eq.~\ref{eq:spatial_sampling} provides a reasonable description of the renewal of $E_\mathrm{b}$ field.

Compared with the renewal mechanisms of previous models for glassy dynamics \cite{cates1997prl,bouchaud2003prl,barrat2018rmp}, 
the most distinct feature of our mechanism is introducing the correlation length $\xi$ that characterizes the medium-range organization of local packing \cite{egami2021pre_1,egami2021pre_2,schweizer2025prl,tanaka2018prx}. 
Note that, structural lengths reflecting medium-range orders in supercooled liquids have been extensively discussed on their role in glass transition \cite{sastry2016ropp, tanaka2018prx,royall2015pr,tanaka2025jpcb, langer2014ropp}, 
while their impact on flow has received much less attention. 
As will be seen later, $\xi$ is indispensable for correctly predicting flow behaviors at both macroscopic and microscopic levels. 

\subsection{Localization of Elasticity}

\label{sec:local_elasticity}

\begin{figure}
	\includegraphics[width=\linewidth]{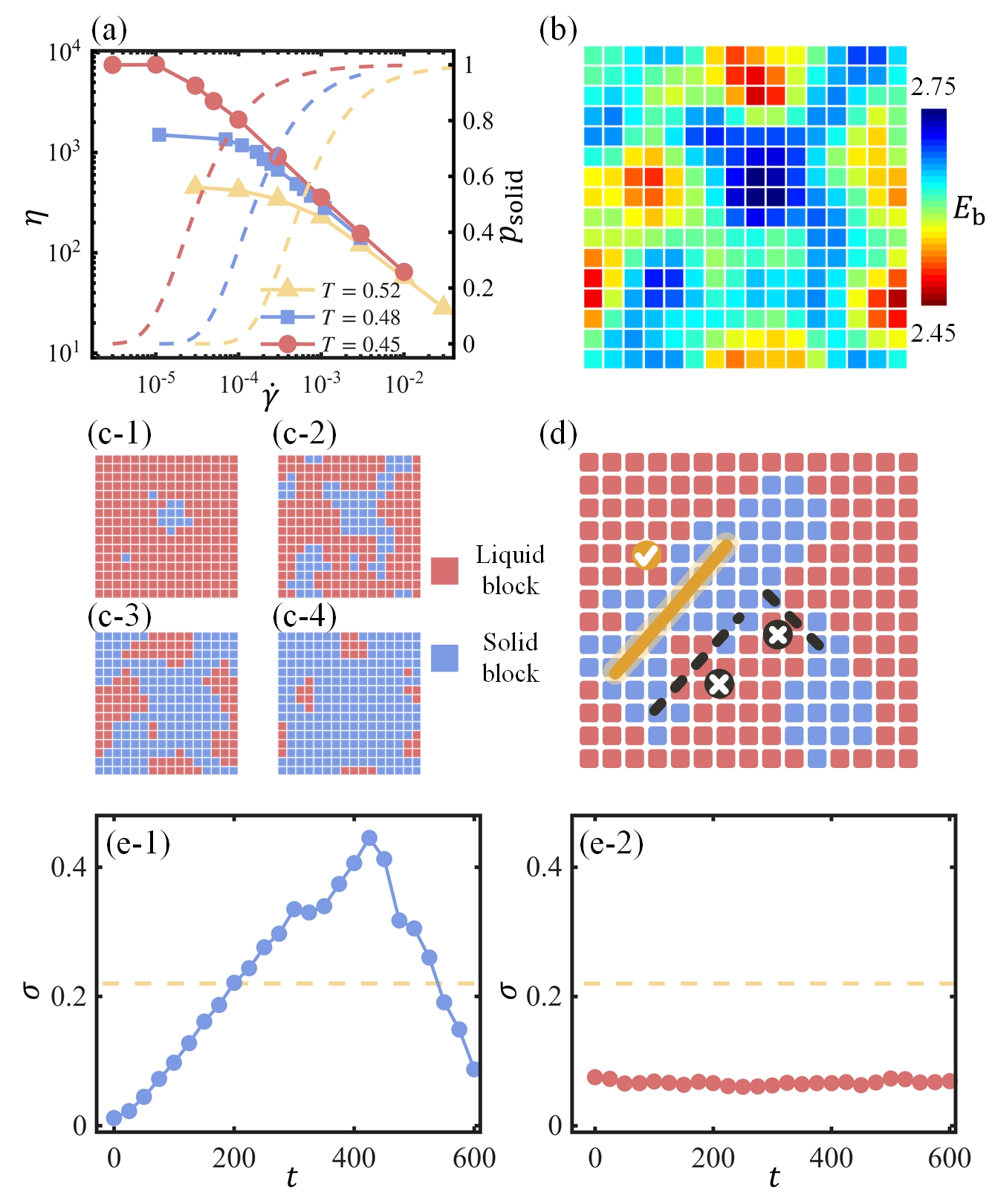}
	\caption{\label{fig:local_elasticity}
		Two-state picture. 
		(a) $\eta$ as a function of $\dot{\gamma}$ for the KA system (solid lines, left axis) 
			and $p_\mathrm{solid}$ as a function of $\dot{\gamma}$ calculated by Eq.~\ref{eq:psolid} (dashed lines, right axis) under steady shear at three temperatures. 
		(b) Representative slice with blocks colored by their barrier height $E_\mathrm{b}$. 
		(c-1)--(c-4) show the corresponding classifications of the blocks in (b) into liquid-like and solid-like states for increasing values of $p_\mathrm{solid}$. 
		(d) Schematic illustration of elastic connectivity: the golden solid line denotes a connected path, whereas the black dashed lines denote disconnected paths. 
		(e) Typical stress variation within a solid-like region (e-1) and that within a liquid-like region (e-2) in our model. 
			In both subpanels, the light-yellow dashed line denotes the system stress. 
	}
\end{figure} 

As $\dot{\gamma}$ grows, more particles become slower than the external rate and, thus, are recruited into solid-like regions, i.e., LERs. 
We denote the fraction of solid-like particles as $p_\mathrm{solid}$. 
An analytical form of $p_\mathrm{solid}(\dot{\gamma})$ can be derived with the shear-facilitated-activation model established in our previous work \cite{wang2025prr}, and is expressed as:
\begin{align}
	\label{eq:psolid}
	p_\mathrm{solid}(\dot{\gamma})= 
	 \exp\Biggl[-\frac{\sqrt{\pi}/2}{\dot{\gamma}\bar{\tau}_\mathrm{LC}\sqrt{\Omega_\mathrm{p}\eta_\mathrm{eq}/(\bar{\tau}_\mathrm{LC} k_\mathrm{B}T)}} \nonumber \\
	 \times \operatorname{erfi}\!\left(\sqrt{\frac{\Omega_\mathrm{p}\eta_\mathrm{eq}}{\bar{\tau}_\mathrm{LC} k_\mathrm{B}T}}\,\gamma_\mathrm{c}\right)\Biggr],
\end{align}
where $\bar{\tau}_\mathrm{LC}$ denotes the particle-averaged $\tau_\mathrm{LC}$ at equilibrium state, 
$\Omega_\mathrm{p}$ is the cage volume, $\eta_\mathrm{eq}$ is the equilibrium viscosity, 
and $\gamma_\mathrm{c}$ is a threshold strain that characterizes whether an activation is mainly driven by thermal effect or shear. 
The detail of the derivation is given in SM. 
Importantly, all parameters in Eq.~\ref{eq:psolid} are obtained from equilibrium state. 
In Fig.~\ref{fig:local_elasticity}(a), we show $p_\mathrm{solid}(\dot{\gamma})$, and replot the MD results of $\eta(\dot{\gamma})$ for comparison. 
In the Newtonian regime, $p_\mathrm{solid}$ is negligible. 
As $\dot{\gamma}$ grows, shear thinning emerges, and $p_\mathrm{solid}$ rapidly increases, which eventually saturates in the power-law-shinning regime. 
Shear thinning can therefore be understood as resulting from the emergence and progressive dominance of the solid-like response. 

From the view of the shear-facilitated-activation picture \cite{wang2025prr}, 
whether the response of a block is solid-like or liquid-like is decisively determined by its undeformed energy barrier, i.e., $E_\mathrm{b}$. 
For a block with large $E_\mathrm{b}$, thermal activation is difficult, 
and there is a great chance that it undergoes a non-negligible affine deformation driven by external shear before its activation, resulting in a solid-like response. 
On the other hand, for a block with small $E_\mathrm{b}$, thermal effect easily activates it before noticeable strain, resulting in a liquid-like response. 
Based on this view, we can introduce a critical barrier $E_\mathrm{c}$ by:
\begin{equation}
	\label{eq:ec}
	\int_{E_\mathrm{c}}^{\infty} \rho(E_\mathrm{b})\,\mathrm{d}E_\mathrm{b} = p_\mathrm{solid}.
\end{equation}
Blocks with $E_\mathrm{b} > E_\mathrm{c}$ are classified as solid-like, otherwise liquid-like. 
Figure~\ref{fig:local_elasticity}(b) and (c) illustrate this classification. 
Figure~\ref{fig:local_elasticity}(b) shows a typical configuration slice in which blocks are colored by their $E_\mathrm{b}$ values. 
Note that the solid-like and the liquid-like blocks are spatially clustered. 
In Fig.~\ref{fig:local_elasticity}(c), we show four cases of spatial partition based on the same configuration shown in Fig.~\ref{fig:local_elasticity}(b) with increasing $\dot{\gamma}$.

In our model, the qualitative difference between solid-like and liquid-like regions is the inter-block interaction. 
Within a solid-like region, the stress of relaxing blocks can be redistributed through the long-ranged anisotropic elasticity-mediated interaction \cite{wang2025jcp}. 
In this case, the change of the stress of block $i$ over a time interval $\mathrm{d}t$ contains a component $\mathrm{d}\sigma_{\mathrm{int},i}$ due to such inter-block stress redistribution: 
$\mathrm{d}\sigma_{\mathrm{int},i}/\mathrm{d}t={\textstyle \sum_{j \ne i} G_{ij}n_j\sigma_j/\tau}$, 
where $G_{ij}$ is the propagator that describes the elasticity-mediated interaction and is commonly set to the Eshelby form \cite{barrat2018rmp,picard2004epje}. 
The situation in liquid-like regions is different. It has been shown that in normal liquids, phonons are strongly damped \cite{egami2013prl}. 
Furthermore, for equilibrium supercooled liquids with temperatures higher than the mode-coupling temperature $T_\mathrm{MCT}$, 
the correlation between flow events is found to be short-ranged \cite{ajliu2021prl}, 
evidently different from the long-ranged, anisotropic character of the Eshelby form.
In our previous work \cite{wang2025jcp,wang2025prr}, we also find that the flow events within a liquid-like region seem highly uncorrelated. 
These results suggest that the long-ranged anisotropic elasticity-mediated interaction is not predominant in liquid-like regions. 
Summarizing the above discussion, we express the block stress increment from the inter-block interaction by:
\begin{equation}
	\label{eq:local_elasticity}
	\frac{\mathrm{d}\sigma_{\mathrm{int},i}}{\mathrm{d}t} = \sum_{j \ne i} C_{ij} G_{ij} \frac{n_j\sigma_j}{\tau},
\end{equation}
where $C_{ij}$ is the elastic connectivity factor. 
It equals $1$ when blocks $i$ and $j$, as well as the blocks crossed by the line connecting blocks $i$ and $j$, are all solid-like; 
otherwise it equals $0$. 
Figure~\ref{fig:local_elasticity}(d) illustrates $C_{ij}$: the golden solid line denotes a connected path ($C_{ij}=1$), whereas the black dashed lines denote disconnected paths ($C_{ij}=0$). 

In typical elastoplastic models (EPM) of amorphous solids, 
the elasticity-mediated interaction spans the entire system \cite{barrat2018rmp}.
Thus, the localization of elasticity represented by $C_{ij}$ constitutes a key difference between flowing supercooled liquids and amorphous solids. 
Figure~\ref{fig:local_elasticity}(e) displays the model results of the mechanical responses of a solid-like region and a liquid-like region. 
The former [Fig.~\ref{fig:local_elasticity}(e-1)] exhibits a significant stress accumulation, 
while the later [Fig.~\ref{fig:local_elasticity}(e-2)] exhibits typical viscous behavior without any noticeable stress accumulation. 
This result, consistent with our previous MD observations \cite{wang2025jcp} (illustrated in Fig.~\ref{fig:ler_review}(b)), 
highlights the fundamental impact of the localization of elasticity on the mechanical behaviors and local dynamics of flowing supercooled liquids. 

\subsection{Two Moduli for Two States}

\label{sec:two_moduli}

It is known that the modulus of glassy materials is spatially inhomogeneous \cite{barrat2013pre,barrat2009pre,yoshimoto2004prl}. 
Due to the heterogeneous distribution of the potential energy barrier and the local response \cite{ajliu2016nat,ciamarra2022prl,wang2025prr}, 
the local modulus of flowing supercooled liquids is also expected to be heterogeneous \cite{lerner2021jcp}. 
Considering the highly-degenerate nature of the two-state picture, we assume two moduli, $\mu_\mathrm{S}$ for solid-like blocks and $\mu_\mathrm{L}$ for liquid-like blocks.

In supercooled liquids, the modulus is transient and depends on the observation time scale \cite{dyre2024jpcl,szamel2015prl}. 
At short time, caging produces a solid-like response and hence a relatively large modulus; 
at longer time, structural relaxation makes the modulus smaller. 
The relevant crossover time scale is the $\alpha$ relaxation time $\tau_\alpha$, which characterizes the time scale of cage breaking. 
For the equilibrium state, the overall modulus can be written as $\mu_\mathrm{eq}=\eta_\mathrm{eq}/\tau_\alpha$ \cite{dyre2024jpcl} 
according to the Maxwell model. 
Considering that all particles are liquid-like at quiescent state, we can use $\mu_\mathrm{eq}$ to represent $\mu_\mathrm{L}$. 
For $\mu_\mathrm{S}$, we set its value to the modulus $\mu_0$ determined by the initial stage of caging, because in this stage the cage still preserves most of elasticity. 
In practice, $\mu_0$ can be extracted from the correlation of transverse particle displacement \cite{keim2015prx, szamel2015prl}, as detailed in SM. 
For the KA system at $T=0.45$, $\mu_0 \approx 13$ and $\mu_\mathrm{eq} \approx 8.5$. 

\subsection{Equation of Motion}

\label{sec:overall_equation}

With above mechanisms and several key ingredients in the EPM model, we write down the stress evolution of block $i$ as:
\begin{equation}
	\label{eq:overall_eq}
	\dot{\sigma}_i
	= \mu_{\mathrm{S,L}}\dot{\gamma}
	+ \sum_{j\ne i} C_{ij} G_{ij}\frac{n_j \sigma_j}{\tau}
	- \frac{n_i \sigma_i}{\tau}.
\end{equation}
The three terms in the right-hand side of Eq.~\ref{eq:overall_eq} respectively represent the external loading with rate $\dot{\gamma}$,
the stress redistribution through inter-block interaction, and the local relaxation. 
Up to now, we obtain a closed constitutive model, in which all parameters are extracted from the equilibrium state. 

\section{Results}

Here, we use the KA system as the reference to examine the effectiveness of the two-state model. 
The examination contains both the macroscopic mechanical response and the microscopic dynamics over a broad range of shear rates. 
Particularly, the roles of the mechanisms proposed in the preceding section will be checked. 

\subsection{Macroscopic Response}

\begin{figure*}
	\includegraphics[width=\linewidth]{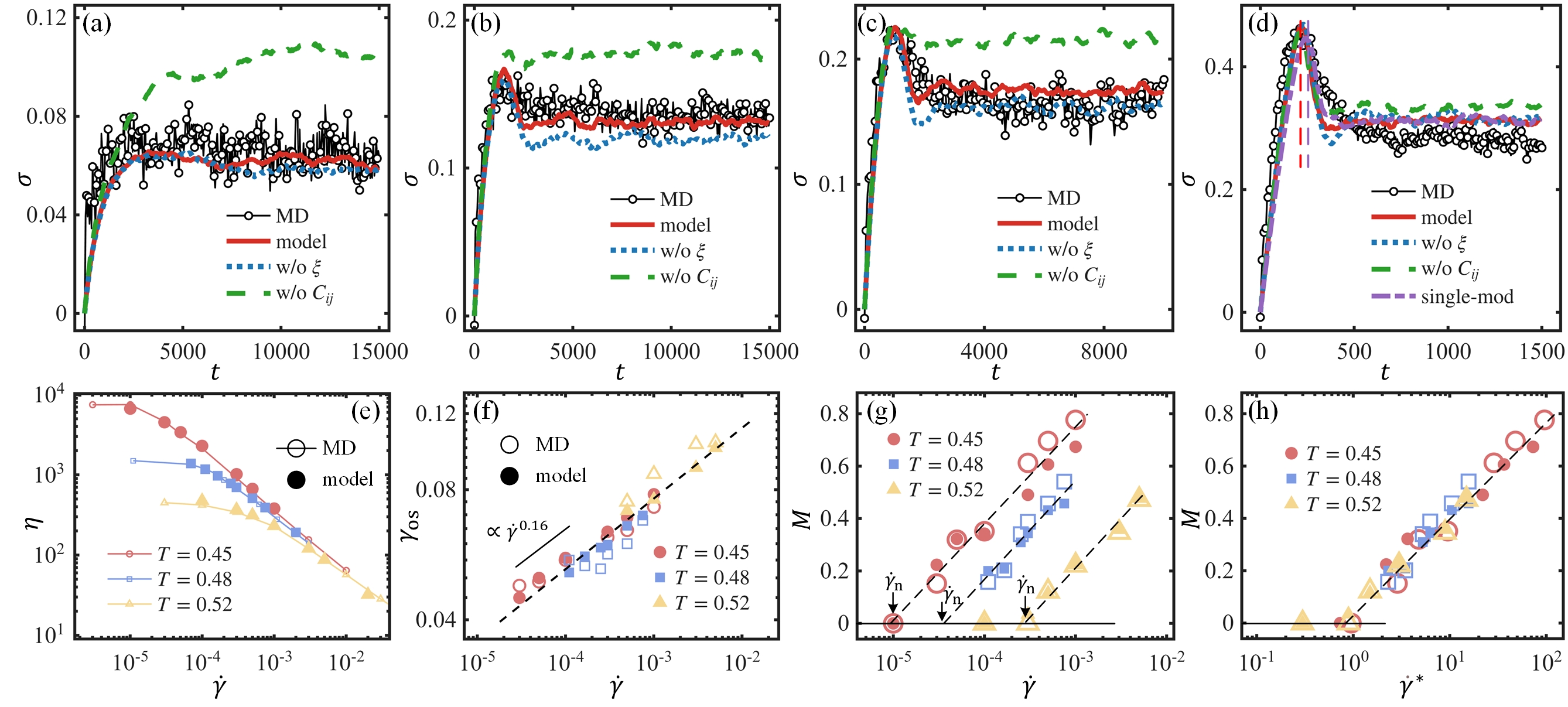}
	\caption{\label{fig:mac_resuluts}
		Mechanical responses of supercooled liquids. 
		(a) -- (d) Start-up stress $\sigma(t)$ at $T=0.45$ for $\dot{\gamma}=10^{-5}$ (a), $3 \times 10^{-5}$ (b), $5 \times 10^{-5}$ (c), and $3 \times 10^{-4}$ (d). 
			In all panels, MD results of the KA system are compared with model predictions. 
			Particularly, predictions from the incomplete models, including a variant without incorporating $\xi$ (denoted as ``w/o $\xi$") 
			and a variant without incorporating localization of elasticity (denoted as ``w/o $C_{ij}$"), are also shown. 
			In (d), we also plot the prediction from the model using only a single modulus $\mu_\mathrm{eq}$. 
			The two vertical dashed lines denote overshoot times of the full model and the single-modulus model. 
		(e) MD results (open symbols) and model predictions (filled symbols) of $\eta$ as a function of $\dot{\gamma}$ under steady shear at three temperatures. 
			The same symbol convention is used in panels (f) -- (h). 
		(f) Overshoot strains predicted by the model and those measured in MD as a function of $\dot{\gamma}$ at three temperatures. 
			The dashed line denotes a power-law fit. 
		(g) Overshoot magnitude $M$ versus $\dot{\gamma}$ for three temperatures. 
			$\dot{\gamma}_\mathrm{n}$ denotes the shear rate where $M$ extrapolates to zero. 
		(h) Overshoot magnitude $M$ versus scaled shear rate $\dot{\gamma}^*=\dot{\gamma}/\dot{\gamma}_\mathrm{n}$ for three temperatures. 
			The dashed line denotes a master curve. 
	}
\end{figure*} 

The transient response to start-up shear provides a sensitive probe of the interplay between intrinsic properties and imposed shear \cite{petekidis2012prl,egelhaaf2017prl,fuchs2013jor}, 
and has been extensively studied for many soft matter systems \cite{polymer_book, dapengbi2022prl, foam, metallic_glass_startup, jiang2015overshoot}. 
Figure~\ref{fig:mac_resuluts}(a) -- (d) show the stress evolutions in start-up shear of the KA system at $T=0.45$ with increasing $\dot{\gamma}$. 
The range of $\dot{\gamma}$ spans the Newtonian regime, the crossover regime and the power-law-thinning regime. 
The model results are also plotted. 
It is seen that the two-state model quantitatively captures the peak strain $\gamma_\mathrm{os}$ and the amplitude of the stress overshoot for all $\dot{\gamma}$. 

The steady-state viscosity can be extracted from the long-time limit of the response to start-up shear. 
Figure~\ref{fig:mac_resuluts}(e) shows both the MD results and the model results of $\eta(\dot{\gamma})$ under steady shear. 
At all studied temperatures, the model accurately predicts the shear thinning.

In Fig.~\ref{fig:mac_resuluts}(a) -- (d), we also plot the results of some ``incomplete" models to check the roles of the mechanisms proposed above. 
Dotted lines represent the model that does not incorporate the correlation length $\xi$ (denoted as ``w/o $\xi$"). 
In this case, the renewal of $E_\mathrm{b}$ is realized only by sampling $\rho(E_\mathrm{b})$. 
Dashed lines represent the model that does not incorporate the localization of elasticity, in other words, $C_{ij}$ is always equal to $1$ (denoted as ``w/o $C_{ij}$"). 
The dash-dot line in Fig.~\ref{fig:mac_resuluts}(d) represents the model that adopts a single modulus $\mu_\mathrm{eq}$ (denoted as “single-mod”). 
When the localization of elasticity is removed, the over-estimation of stress is most pronounced at low $\dot{\gamma}$, 
where the system remains predominantly liquid-like and the global elasticity is physically inappropriate. 
The agreement only improves at large $\dot{\gamma}$, where the fraction of solid-like blocks becomes dominant. 
When the correlation length $\xi$ is removed, the model under-estimates the stress. 
The deviation is most evident when $\dot{\gamma}$ is in the crossover regime (Fig.~\ref{fig:mac_resuluts}(b)), 
where the fractions of the two states are comparable and the spatial organization of these states has the largest effect on the macroscopic response. 
Figure~\ref{fig:mac_resuluts}(d) examines the effect of the dual modulus. 
It is seen that by using a single modulus, the predicted $\gamma_\mathrm{os}$ becomes larger than the observed one. 
Such difference is enhanced by increasing $\dot{\gamma}$, suggesting the importance of the progressively growing solid-like component when the system is driven into deeper nonlinear regime. 

We now turn to the characteristics of the stress overshoot itself. 
Figure~\ref{fig:mac_resuluts}(f) shows the overshoot strain $\gamma_\mathrm{os}$ as a function of $\dot{\gamma}$ at different temperatures. 
Both MD and model give a power law $\gamma_\mathrm{os} \sim \dot{\gamma}^\kappa$ with $\kappa \approx 0.16$ for all temperatures. 
Moreover, we extract the data from an experiment on hard-sphere colloids whose concentrations are close to the glass transition point \cite{petekidis2012prl}. 
The experimental result gives $\gamma_\mathrm{os} \sim \dot{\gamma}^{0.18}$, close to the MD and model results. 

The magnitude of the overshoot can be characterized by $M=\sigma_\mathrm{pk}/\sigma_\mathrm{steady} - 1$, 
where $\sigma_\mathrm{pk}$ is the maximum stress of the overshoot and $\sigma_\mathrm{steady}$ is the steady stress. 
Figure~\ref{fig:mac_resuluts}(g) shows $M(\dot{\gamma})$ at different temperatures. 
Here, model results well agree with the MD results. 
When $\dot{\gamma}$ is small, the stress overshoot is negligible \cite{fuchs2013jor}. 
Once $\dot{\gamma}$ exceeds a certain shear rate $\dot{\gamma}_\mathrm{n}$, overshoot emerges and $M$ increases with $\ln \dot{\gamma}$ in a linear way, as shown in Fig.~\ref{fig:mac_resuluts}(g). 
$\dot{\gamma}_\mathrm{n}$ coincides the end of the Newtonian regime in the steady viscosity curve $\eta(\dot{\gamma})$, 
linking the emergence of stress overshoot to the onset of nonlinear rheology. 

$M(\dot{\gamma})$ curves at different temperatures collapse when $\dot{\gamma}$ is rescaled by the corresponding $\dot{\gamma}_\mathrm{n}$, 
$\dot{\gamma}^*=\dot{\gamma}/\dot{\gamma}_\mathrm{n}$, as shown in Fig.~\ref{fig:mac_resuluts}(h). 
This collapse indicates that the overshoot magnitude is controlled primarily by the ratio $\dot{\gamma}/\dot{\gamma}_\mathrm{n}$, 
rather than by temperature separately, over the range studied here. 
This prediction is directly testable in future experiments and simulations.

\subsection{Microscopic Dynamics}

\label{sec:mic_results}

\begin{figure}
	\includegraphics[width=\linewidth]{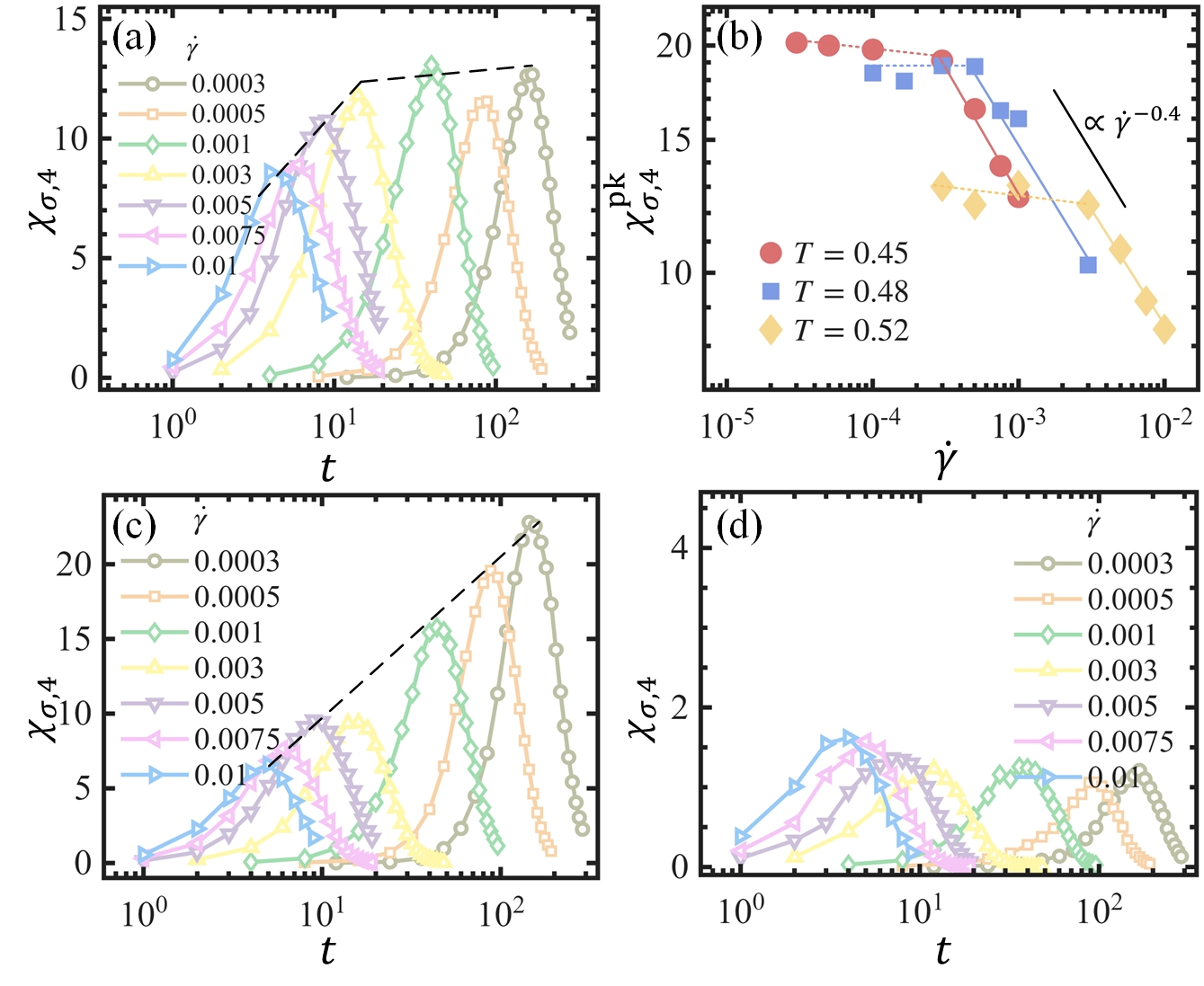}
	\caption{\label{fig:mic_results}
		Microscopic DH in flowing supercooled liquids predicted by model. 
		(a) Four-point susceptibility $\chi_{\sigma, 4}(t)$ under different $\dot{\gamma}$ at $T=0.52$. 
		(b) Peak susceptibility $\chi_{\sigma, 4}^\mathrm{pk}$ as a function of $\dot{\gamma}$ for three temperatures. 
			Solid lines denote power-law fits in the high-$\dot{\gamma}$ regime. 
		(c) $\chi_{\sigma, 4}(t)$ at $T=0.52$ predicted by an incomplete model with the localization of elasticity being removed (the ``w/o $C_{ij}$" variant). 
		(d) $\chi_{\sigma, 4}(t)$ at $T=0.52$ predicted by another incomplete model with the correlation length $\xi$ being removed (the ``w/o $\xi$" variant). 
	}
\end{figure} 

To characterize DH in the model, we track the spatial distribution of the cumulative stress drop $\sigma_\mathrm{dn}$ over a given time interval. 
For block $i$, $\sigma_{\mathrm{dn},i}(t)$ records the total stress released in relaxation events during the time interval $t$. 
Large values of $\sigma_\mathrm{dn}$ identify regions that have undergone substantial local relaxation and are therefore likely to promote nearby relaxation events. 
We quantify the resulting heterogeneity through a four-point susceptibility \cite{szamel2014prl,castillo2023prl}, 
\begin{equation}
	\label{eq:mic_dh_x4}
	\chi_{\sigma, 4}(t)=\lim_{q\to 0}\left\langle
	\frac{1}{N}\sum_{i,j}\delta h_i(t)\,\delta h_j(t)\cos\!\left(\vec{q}\cdot\vec{r}_{ij}\right)
	\right\rangle,
\end{equation}
where $\delta h_i=h_i-N^{-1}\sum_j h_j$, and $h_i$ is given by
\begin{equation}
	\label{eq:mic_dh_hit}
	h_i(t)=H\!\left[\sigma_{\mathrm{dn},i}(t)-\sigma_\mathrm{c}\right],
\end{equation}
where the threshold stress $\sigma_\mathrm{c}$ is determined from the distribution of $\sigma_\mathrm{dn}$, as described in SM. 
The results reported below are not sensitive to the precise value of $\sigma_\mathrm{c}$.

Figure~\ref{fig:mic_results}(a) shows $\chi_{\sigma, 4}(t)$ predicted by the two-state model for the $T=0.52$ KA sample at different $\dot{\gamma}$. 
As $\dot{\gamma}$ increases, the peak position of $\chi_{\sigma, 4}(t)$ shifts to shorter time. 
Meanwhile, the peak height $\chi_{\sigma, 4}^\mathrm{pk}$ exhibits a crossover. 
Figure~\ref{fig:mic_results}(b) shows $\chi_{\sigma, 4}^\mathrm{pk}$ as a function of $\dot{\gamma}$. 
The crossover to a power-law shrinkage $\chi_{\sigma, 4}^\mathrm{pk} \sim \dot{\gamma}^{-\theta}$ with $\theta \approx 0.4$ is clearly seen for all temperatures. 
The crossover shear rate is close to $\dot{\gamma}_\mathrm{pl}$, where the power-law thinning appears in the steady viscosity. 
The model results shown in Fig.~\ref{fig:mic_results}(a) and (b) are highly consistent with the MD results shown in Fig.~\ref{fig:md_info}(c) and (d). 

The microscopic origin of the crossover in $\chi_{\sigma, 4}^\mathrm{pk}(\dot{\gamma})$ can be explored by the ``incomplete" models. 
The $\chi_{\sigma, 4}(t)$ calculated by the model that removes the localization of elasticity (the ``w/o $C_{ij}$" model) are given in Fig.~\ref{fig:mic_results}(c). 
In this case, the crossover disappears. 
The $\chi_{\sigma, 4}(t)$ calculated by the model that removes the correlation length $\xi$ (the ``w/o $\xi$" model) are given in Fig.~\ref{fig:mic_results}(d). 
In this case, the magnitude of $\chi_{\sigma, 4}^\mathrm{pk}$ becomes weaker by one order of magnitude, and its $\dot{\gamma}$-dependence is also suppressed. 
These two tests show that the crossover behavior arises from the interplay between barrier clustering and localized elasticity.
At low $\dot{\gamma}$, the large fraction of liquid-like regions interrupts elastic connectivity and renders $\chi_{\sigma, 4}^\mathrm{pk}$ insensitive to $\dot{\gamma}$. 
At higher $\dot{\gamma}$, the solid-like fraction becomes dominant, and the flow is governed by LERs.
LER shrinks with $\dot{\gamma}$ at high-$\dot{\gamma}$ regime \cite{wang2022prx}, resulting in the shrinkage of dynamic length \cite{wang2025jcp}.

\section{Discussion and Concluding Remarks}

Supercooled liquids occupy an ``uncomfortable" position between normal liquids and amorphous solids, making the quantitative understanding of their flow behaviors more lagging. 
On the one hand, the two-point framework developed for normal liquids \cite{hansen2013book,hess1983pla,hess1986pra,hess1987pra,yip1983pra,dhont1996book} seems difficult for adequately 
incorporating supercooled liquids' hallmarks such as the heterogeneities in dynamics and structure and the facilitation in dynamics. 
On the other hand, models for the deformation of amorphous solids focus on the solid-to-liquid transition 
during yielding \cite{bonn2017rmp,langer2011review,berthier2025natrev,barrat2018rmp,wangweihua2012pms,biroli2022prl,fielding2020prl,martens2024prl,zaccone2021prl}, 
particularly at the limits of low flow rate and low temperature \cite{maloney2006pre,martens2021prl, martens2016prl, wyart2014pnas}, 
which is not the key problem of supercooled liquids. 
These models heavily rely on the picture of ``soft regions" that yield within a rigid background \cite{langer1998pre, ajliu2014prx, manning2020prm,zaccone2022pre}. 
For supercooled liquids, however, the interplay between dynamic heterogeneity and finite shear rate calls for a different perspective. 
Regions that relax more slowly than the imposed shear are converted into LERs that behave as localized amorphous solids. 
The relevant objects are therefore not ``soft regions" in a rigid solid, but ``hard regions" embedded in a flowing liquid. 
The two-state model makes this picture explicit. 

In principle, energy levels and relaxation pathways of local structure in supercooled liquids are very rich. 
Coarse graining merges these microscopic variations into two effective states while preserving their spatial organization. 
This is precisely what the field of clustered energy barrier $E_\mathrm{b}$ encodes. 
Similar logic underlies other two-state models for amorphous systems \cite{langer2013pre,tanaka2010natmat,tanaka2012epje,langer1998pre,ginzburg2020sofmat}. 
A noticeable characteristic of our model is the introduction of the structural correlation length $\xi$, which reflects the interaction between regions in block renewal. 
$\xi$ is obtained from equilibrium and set to be $\dot{\gamma}$-independent. 
By contrast, the dynamic correlation length, which can be characterized by $\chi_4^\mathrm{pk}$ \cite{glotzer2003jcp,yamamoto2012jcp}, shrinks as $\dot{\gamma}$ increases. 
This decoupling between structural and dynamic lengths stands in contrast to the glass transition of quiescent states, 
where recent studies reported coherent growth of structural and dynamic lengths upon cooling 
\cite{tanaka2018prx,tanaka2010natmat,Bhattacharyya2025jcp,tanaka2025jpcb,schall2024pnas,sastry2018prl,sastry2017prl,ilg2010prl}. 
A possible interpretation appeals to how dynamic events develop in space. 
According to Ref.~\cite{tanaka2018prx}, microscopic structure provides a ``template" that the dynamic field gradually fills as relaxation extends from local to longer ranges over time. 
The template sets the upper bound that the dynamic length can reach in equilibrium, and only temperature reshapes the template itself. 
Whereas under shear conditions, dynamic events are influenced by both external drive and structure. 
Shear interrupts the buildup of the dynamic field upon the template, contracting the dynamic length while leaving the structural length intact. 
With this picture, our model reveals the role of structural correlation length in shaping the response to external deformation. 

In a conventional view, shear effect can be embodied in the effective temperature framework \cite{barrat2002jcp,xvning2005prl}, 
which treats shear as a source of effective heating that enhances diffusion and gives rise to shear thinning. 
Our picture suggests a contrasting picture: shear does not make the system more fluid, but rather partly more solid. 
Shear, then, does not eliminate but rather highlights the glassy features of a supercooled liquid. 
This perspective could be related to the long-standing problem of the glass transition of supercooled liquids, 
where competing views have emphasized different mechanisms for the macroscopic slowdown and microscopic heterogeneity that develop upon cooling. 
Some highlight dynamic facilitation \cite{chandler2011prx}, others structural orders \cite{tanaka2025jpcb,royall2015pr}, 
and still others growing elastic correlations \cite{dyre2024jpcl,dyre2006rmp}. 
The absence of a consensus may indicate that each of these views capture a facet of the same underlying physics. 
Our results offer an opportunity to understand this problem from flowing states. 
Within an LER, elasticity-mediated cooperativity drives correlated rearrangements, providing a mesoscopic substrate for dynamic facilitation. 
More specifically, a shear transformation zone within an LER triggers further rearrangements through an anisotropic elastic field \cite{wang2025jcp}, 
a mechanism that has also been argued to play an important role in equilibrium supercooled liquids below the mode-coupling temperature \cite{ajliu2021prl}. 
Moreover, here we reveal an intimate connection between medium-range structural order and flow behaviors, consistent with structural perspectives. 
Thus, we suggest that LERs offer an indirect view for glass-transition problems. 
Recent observations support this idea: 
It is suggested that equilibrium relaxations bear striking similarity to those under shear \cite{patinet2022prl, ajliu2021prl}. 
These findings indicate a deep connection between the equilibrium and driven states of supercooled liquids.

The present modeling is applied to the KA system, and extending the framework to other glass-forming systems will be an important test of its generality. 
Beyond this, several promising extensions of the current model are within reach. 
Beyond steady shear, our model may be generalized to small/large amplitude oscillatory shear \cite{rogers2021prl, rogers2025prl_1, rogers2025prl_2}, 
where the interplay between reversible elasticity and irreversible relaxation should provide a stringent test of the two-state picture. 
Surface flow in supercooled liquids \cite{voigtmann2026prl} is another promising direction, since the spatial correlation length is likely to play an even more prominent role near interfaces. 
More broadly, recent progress in adapting EPMs to equilibrium states \cite{biroli2023prl, biroli2023prx, ajliu2024arxiv} and active matter \cite{sollich2025sofmat, manning2025arxiv} suggests that 
analogous extensions of our model may provide a useful route toward a broader mesoscopic description of disordered systems.

\begin{acknowledgments}
This research was partially supported by National Natural Science Foundation of China (no. 11975136). 
Computational resources were provided by the Center of High Performance Computing, Tsinghua University. 
\end{acknowledgments}

\appendix

\section{simulation}

\label{sec:appendix_sim}

In the MD simulation, we use the Kob-Andersen binary mixture \cite{kobandersenmixture} as the model system. 
It is a binary mixture of particles (type A: 80\%, type B: 20\%). 
Particles interact via a Lennard-Jones (LJ) potential, 
$V(r)=4\epsilon_{\alpha\beta}\left[\left(\sigma_{\alpha\beta}/r\right)^{12}-\left(\sigma_{\alpha\beta}/r\right)^{6}\right]$, 
where $r$ is the interparticle distance and $\alpha,\beta\in{\mathrm{A},\mathrm{B}}$ denote particle types. 
The interaction parameters are $\epsilon_{\mathrm{AA}}=1.0$, $\epsilon_{\mathrm{AB}}=1.5$, $\epsilon_{\mathrm{BB}}=0.5$,
and $\sigma_{\mathrm{AA}}=1.0$, $\sigma_{\mathrm{AB}}=0.8$, $\sigma_{\mathrm{BB}}=0.88$. 
All particles have identical mass $m$. 
We employ reduced Lennard-Jones units: length in unit of $\sigma_{\mathrm{AA}}$, energy in $\epsilon_{\mathrm{AA}}$, time in $\sigma_{\mathrm{AA}}\sqrt{m/\epsilon_{\mathrm{AA}}}$, 
temperature in $\epsilon_{\mathrm{AA}}/k_{\mathrm{B}}$, and stress in $\epsilon_{\mathrm{AA}}/\sigma_{\mathrm{AA}}^3$. 
The potential is truncated and shifted at $r_\mathrm{cut}=2.5$. 
Equations of motion are integrated with a time step $\mathrm{d}t=0.005$ in the canonical (\textit{NVT}) ensemble using a Nose-Hoover thermostat. 
The number density is fixed at $\rho=1.2$. 
Simulations are performed in a three-dimensional cubic box with periodic boundary conditions, containing $N=108,000$ particles. 
We investigate three temperatures, $T=0.52$, $0.48$, and $0.45$, all within the supercooled regime. 
Steady shear is imposed using the SLLOD equations of motion combined with Lees-Edwards boundary conditions \cite{lammps}.

\bibliography{mainreferences}

\end{document}